\input harvmac
\input psfig
\newcount\figno
\figno=0
\def\fig#1#2#3{
\par\begingroup\parindent=0pt\leftskip=1cm\rightskip=1cm\parindent=0pt
\global\advance\figno by 1
\midinsert
\epsfxsize=#3
\centerline{\epsfbox{#2}}
\vskip 12pt
{\bf Fig. \the\figno:} #1\par
\endinsert\endgroup\par
}
\def\figlabel#1{\xdef#1{\the\figno}}
\def\encadremath#1{\vbox{\hrule\hbox{\vrule\kern8pt\vbox{\kern8pt
\hbox{$\displaystyle #1$}\kern8pt}
\kern8pt\vrule}\hrule}}
\def\underarrow#1{\vbox{\ialign{##\crcr$\hfil\displaystyle
 {#1}\hfil$\crcr\noalign{\kern1pt\nointerlineskip}$\longrightarrow$\crcr}}}
%
\def\hat{\widehat}
\def\tilde{\widetilde}
\overfullrule=0pt
\def\C{{\bf C}}
\def\IC{{\bf C}}
%
\def\tilde{\widetilde}
\def\bar{\overline}
\def\Z{{\bf Z}}
\def\T{{\bf T}}
\def\RP{{\bf RP}}
\def\S{{\bf S}}
\def\R{{\bf R}}

\font\zfont = cmss10 
\font\litfont = cmr6

\def\bigone{\hbox{1\kern -.23em {\rm l}}}
\def\ZZ{\hbox{\zfont Z\kern-.4emZ}}
\def\half{{\litfont {1 \over 2}}}

\Title{hep-th/9912279}
{\vbox{\centerline{Self-Duality, Ramond-Ramond Fields,}
\bigskip
\centerline{and K-Theory   }}}
\smallskip
\centerline{Gregory Moore$^*$ }
\smallskip
\centerline{\it Department of Physics, Yale University, New Haven CT
06520 USA}
\smallskip
\smallskip
\smallskip
\centerline{Edward Witten$^*{}^*$}
\centerline{\it Department of Physics, Caltech, Pasadena, CA 91125}
\centerline{\it and   }
\centerline{\it Caltech-USC Center for Theoretical Physics, Univ. of
Southern California}\bigskip
\vskip 1cm
\noindent
\noindent
Just as $D$-brane charge of Type IIA and Type IIB superstrings is
classified, respectively, by $K^1(X)$ and $K(X)$, Ramond-Ramond fields
in these theories are classified, respectively, by $K(X)$ and $K^1(X)$.
By analyzing a recent proposal for how to interpret quantum self-duality
of RR fields, we show that the Dirac quantization formula for the 
RR $p$-forms, when properly formulated, receives corrections that reflect 
curvature, lower brane charges, and an  anomaly of $D$-brane world-volume
fermions.  The $K$-theory framework is important here, because the
term involving the fermion anomaly cannot be naturally
expressed in terms of cohomology and differential forms.

\vskip 1cm
\noindent
$^*$ Address after January 1, 2000: Department of Physics, Rutgers
University, Piscataway NJ 08855-0849 USA.

\noindent
$^*{}^*$ On leave from Institute for Advanced Study, Princeton NJ 08540 USA.
\medskip

\Date{December, 1999}

\newsec{Introduction}

The Ramond-Ramond (RR) fields
of Type IIA superstring theory are differential
forms $G_0,G_2,G_4,\dots$ of all even orders, while for Type IIB superstring
theory, one has RR fields $G_1,G_3,G_5,\dots$ of all odd orders.
The total RR field $G=G_0+G_2+G_4+\dots$ or $G=G_1+G_3+G_5+\dots$
is, classically, self-dual.  

Self-duality alone introduces a number of subtleties in the study of the
RR fields.  For example, one expects a Dirac quantization condition for the
$G_p$'s, naively
\eqn\tolgo{\int_{U_p}{G_p\over 2\pi}\in \Z,}
for every $p$-cycle $U_p$ in spacetime.  We will see that this statement
receives modifications from several sources.  Self-duality makes
the interpretation of any such statement delicate, 
since classically one cannot
impose the relation \tolgo\ for $G_p$ and the dual field $G_{10-p}=*G_p$
at the same time; this point is explained in section 3 of 
\ref\witten{E. Witten,
``Duality Relations Among Topological Effects In String Theory,''  hep-th/9912086.}.

The effects of self-duality are most obvious for the five-form $G_5$
of  Type IIB.  $G_5$ is self-dual, and this makes its dynamics
particularly subtle, as has long
been appreciated \ref\marcusschwarz{N. Marcus and J. H. Schwarz, ``Field
Theories That Have No Manifestly Lorentz Invariant Formulation,''
Phys. Lett. {\bf B115} (1982) 111.}.
One approach to the quantum mechanics of such a field is to construct
its quantum partition function by identifying the right theta function,
as suggested in \ref\oldwitten{E. Witten, ``Fivebrane Effective Action
In $M$-Theory,'' J. Geom. Phys. {\bf 22} (1997) 103, hep-th/9610234.}
\nref\hns{M. Henningson, B. E. W. Nilsson, and P. Salomonson, ``Holomorphic
Factorization Of Correlation Functions In $(4k+2$)-Dimensional $2k$-Form
Gauge Theory,'' hep-th/9908107.}%
and developed more explicitly in \refs{\hns}.  (References to a variety
of other approaches to self-dual dynamics can be found in \witten.)
Except for this one case, one might hope at first sight to eliminate
the subtleties of self-duality by eliminating the $G_p$ of $p>5$ using
$G=*G$ and
treating the $G_p$ of $p<5$ (or rather, their potentials) as the independent
variables.  

\nref\mm{G. Moore and R. Minasian,  ``$K$ Theory And Ramond-Ramond Charge,''
JHEP {\bf 9711:002} (1997), hep-th/9710230.}%
\nref\ew{E. Witten, ``$D$-Branes And $K$ Theory,'' JHEP {\bf 812:019} (1998),
ehp-th/9810188.}%
But things are not so simple.  Eliminating
the $G_p$'s of $p>5$  is unnatural  since it is not invariant under $T$-duality.
Moreover, the $G_p$'s of different $p$ are coupled in a subtle
way  by the reinterpretation  \refs{\mm,\ew}
of RR charges and (as we will argue)
RR fields in $K$-theory. This again makes it subtle to eliminate
half of them.  

The goal of the present paper is to explain what statements along
the line of \tolgo\ mean and what they say given the self-duality of RR
fields and their interpretation in $K$-theory.  For this we need
first of all a precise framework for how to interpret RR fields
(as opposed to charges) in $K$-theory.  This is the subject of section
2.  In brief, we propose that the cohomology class of
the  RR field of Type IIA superstring
theory in a spacetime $X$ is determined by
an element of $K(X)$, while for Type IIB
it is determined by
an element of $K^1(X)$.  (The relation is stated in eqn. (2.17) below.)
This is precisely the opposite relation from
the charges; we recall that RR charge of Type IIA takes values in $K^1(X)$
(with a compact support condition),
while for Type IIB it takes values in $K(X)$.  In section 3, we discuss
how we implement self-duality in $K$-theory.  In this, we follow the
framework described in sections 3 and 4.3 of \witten\ (which are recommended
as background to the present paper); we summarize and amplify 
some key points.  In particular, the mod 2 index of a certain Dirac
operator plays an important role in this discussion.
Finally in section 4, we apply this framework to the quantization conditions
obeyed by the $G_p$'s  for Type IIA.

For the special case of $G_4$, the shifted quantization law
we get has been obtained before by considering global anomalies of membranes
\ref\ugwitten{E. Witten, ``On Flux Quantization In $M$ Theory And
The Effective Action,'' J. 
Geom. Phys. {\bf 22} (1997) 1, hep-th/9609122.}.  
(That derivation was  formulated
for $M$-theory but applies equally in Type IIA.)  The shifts we get
in general are natural from the standpoint of brane anomalies, even
though we include no branes in the derivation.  The shifted quantization
laws cannot in general be naturally stated in cohomology, because  there
is in general no cohomological formula for the fermion  global anomaly
that enters in the analysis.
This is another reason that it is important to describe RR fields via
$K$-theory instead of cohomology.  

Indeed, finding a natural framework  in which to formulate
the shifted quantization condition of $G_6$ was the original goal of the 
present paper.  We started with the anomaly cancellation
condition (4.4) and hoped to use it to understand just what kind of
objects RR fields are.  This proved difficult
because it was hard to understand the role of the fermion anomaly.
 Following progress on understanding self-duality of RR fields in 
$K$-theory \witten, it became clear, as we show in section 4, that
in that framework the fermion anomaly term in the quantization condition 
comes in automatically.  An important role in the $K$-theory framework
is played by eqn. (2.17), which we motivate in section 2 in 
a fairly  elementary
way using the same brane couplings that lead to eqn. (4.4).
This formula was suggested by D. Freed as an interpretation of eqn. (4.4)
with the fermion terms dropped.

In the present paper, the NS three-form field $H$ is assumed to vanish.
Including it raises a number of new issues, some of which will hopefully
be addressed elsewhere \ref\dmw{E. Diaconescu, G. Moore, and E. Witten,
to appear.}.

\newsec{RR Fields And $K$-Theory}

Before considering Type II superstring 
theory, where the RR fields should really be
interpreted in $K$-theory, let us consider the theory of an ordinary
$(p-1)$-form potential $C_{p-1}$, with field strength $G_p$, in 
a $(d+1)$-dimensional spacetime $X$. (In superstring theory, $d=9$.)
We will work first on a spacetime
$X=\R\times M$, where $\R$ is the ``time'' direction and $M$ is the
spatial manifold. We assume
that there are branes in $M$ of codimension $p+1$ 
that serve as magnetic
sources for $C_{p-1}$.  (By replacing $G_p$ with $*G_p$ in the following,
we could similarly consider electric sources.)
 In the presence of such a brane with worldvolume $W$, $G_p$ obeys
\eqn\loxxo{dG_p=\delta(W),}
where $\delta(W)$ is a $(p+1)$-form delta function that is Poincar\'e dual
to $W$.   Here we are writing this formula in the simple form that would
hold for ordinary $p$-form fields (as opposed to the RR fields of Type II,
for which there are additional terms whose import is discussed below).

In general, for any brane worldvolume $W$ in $M$,
$\delta(W)$ is a closed $(p+1)$-form that defines an element $[W]\in
H^{p+1}(M;\Z)$.  
Thus, we interpret the brane charge in this situation as an element of
this cohomology group.
\loxxo\ says that $[W]$ is of the form $d(\dots)$,
\foot{We assume that the equation \loxxo, which is stated for differential
forms, is an approximation to an equation that holds for the integral
cohomology.}
 so that the cohomology
class that represents the brane charge is zero (or else the equation for
$G_p$          has no solution).

If $M$ is a compact manifold without boundary, this is the right answer:
the total brane charge is identically zero.
The way this is often stated is that the total charge associated
with an abelian gauge symmetry vanishes on a compact manifold, since
``the flux has nowhere to go.''  

For a setting in which the total brane charge is not
zero, we consider the case that $M$ is 
noncompact with ``boundary''
$ N$.  We  use the term ``boundary''
 somewhat loosely; a typical case of interest is
that $M=\R^d$, and $N$ is the sphere $\S^{d-1}$ at infinity.  

Even when $M$ is not compact,
\loxxo\ still implies that $[W]$ vanishes as an element of $H^{p+1}(M;\Z)$.
However, if we assume that $W$ is compact, we can consider $[W]$ as an 
element of the compactly supported cohomology $H_{cpct}^{p+1}(M;\Z)$.
\loxxo\ says that the class of $[W]$ vanishes in $H^{p+1}(M;\Z)$,
but it does not imply that $[W] $ vanishes in $H^{p+1}_{cpct}(M;\Z)$.
The reason for this is that even if a $G$-field obeying \loxxo\ exists,
it may not vanish at infinity.  Vanishing of $[W]$ as an element of
$H^{p+1}_{cpct}(M;\Z)$ would imply that there exists a solution $G$ of
\loxxo\ that vanishes at infinity.

Thus, we should regard the brane charge as an element of $H^{p+1}_{cpct}(M;\Z)$
that vanishes if mapped to $H^{p+1}(M;\Z)$.  The brane charge, in other
words, takes values in the kernel of the natural map
\eqn\ucni{i:H^{p+1}_{cpct}(M;\Z)\to H^{p+1}(M;\Z)}
which is defined by ``forgetting'' that a cohomology class has compact support.

So far we have tried to define the brane charge                               
directly in terms of the brane worldvolume.
However, it is often very useful
in gauge theories to define the charge in terms of the behavior of the fields
at infinity -- or in this case, in terms of the restriction of $G_p$ to
$N$.  Indeed, we have seen two paragraphs ago that the brane charge
vanishes if $G_p$ vanishes when restricted to $N$, strongly suggesting
that the brane charge can be measured from the restriction of $G_p$ to $N$.

We can get such a description of the brane charge using the
long exact cohomology sequence for the pair $(M,N)$.  It reads
\eqn\ploo{\dots H^p(M;\Z)\underarrow{j} H^p(N;\Z) \to
H^{p+1}(M,N;\Z)\underarrow{i}H^{p+1}(M;\Z)\to\dots.}
Here the relative cohomology $H^{p+1}(M,N;\Z)$ is the same as the
cohomology with compact support $H^{p+1}_{cpct}(M;\Z)$.
The map $j$ is defined by restricting a cohomology class of $M$ to the
boundary $N$.  From \ploo, we learn that
\eqn\kro{{\rm ker}(i)=H^p(N;\Z)/j(H^p(M;\Z)).}
This shows that the brane charge is determined by the cohomology class
of the $G$-field in $H^p(N;\Z)$, but that elements of $H^p(N;\Z)$ that
arise by restricting to $N$ a cohomology class on $M$ should be considered
to represent zero brane charge.
This has a simple intuitive interpretation.  A $G$-field
on $N$ that extends over $M$ as a closed $p$-form has no brane source
and so has not been ``created'' by branes.  
Such a $G$-field is measureable on $N$ but does not contribute
to the brane charge, which takes values in the quotient indicated
in \kro.

\bigskip\noindent{\it Analog In $K$-Theory}

\nref\vafa{M. Bershadsky, C. Vafa, and V. Sadov, ``$D$-Branes And
Topological Field Theories,'' Nucl. Phys. {\bf B463} (1996) 420.}%
\nref\ghm{M. B. Green, J. A. Harvey, and G. Moore, ``$I$-Brane Inflow
And Anomalous Couplings On $D$-Branes,'' Class. Quant. Grav. 
{\bf 14} (1997) 47, hep-th/9605033.}%
\nref\chyi{E. Cheung and Z. Yin, ``Anomalies, Branes, and Currents,''
Nucl. Phys. {\bf B517} (1998) 69, hep-th/9710206.}%
Now we move on to the Type II case, which differs in a few ways.
There are $G$-fields
of all even or all odd $p$, the brane $W$ supports a Chan-Paton gauge
bundle, and there are a number of subtle corrections to \loxxo\
involving lower brane charges \refs{\vafa,\ghm,\chyi,\mm}.

Thus, considering first Type IIB, for branes of compact support in space,
the brane charge is an element of $K_{cpct}(M)$, the compactly
supported $K$-theory of $M$.  To ensure that the equation for the RR
fields has a solution, the brane charge must
 map to zero in $K(M)$.  Thus, the brane charge
takes values in the kernel of the natural map
\eqn\kkix{i:K_{cpct}(M)\to K(M)}
which ``forgets'' that a $K$-theory class has compact support.

Just as in the case that the brane charge is interpreted as a cohomology
class, we also want a description in which the brane charge is measured
in terms of the RR fields at infinity.  To see what form this must take,
we look at the exact sequence that is the $K$-theory counterpart of
\ploo.  It reads
\eqn\julpo{\dots \to K^{-1}(M)\underarrow{j} K^{-1}(N)
\to K(M,N)\underarrow{i} K(M)\dots,}
where again the relative $K$-group $K(M,N)$ is the same as
$K_{cpct}(M)$, and $j$ is the map that restricts a $K$-theory class
on $M$ to $N$.  (This exact sequence has been used in
computing $K_{cpct}(M)$ \ref\gukov{S. Gukov, ``$K$ Theory, Reality,
and Orientifolds,'' hep-th/9901042.}.)
By the periodicity theorem, $K^{-1}$ is the same as $K^1$.  We
see that the group ${\rm ker}(i)$ in which  the brane charge takes
values has an alternative description:
\eqn\pgo{{\rm ker}(i)= K^1(N)/j(K^1(M)).}
We interpret this the same way that we did in the case of cohomology.
$K^1(N)$ classifies RR fields at infinity, while $K^1(M)$ classifies
RR fields on $M$ that do not have any brane sources.  An RR field on $N$ that
extends (as an element of $K^1$) over $M$ does not require any brane
sources, so the brane charges are classified by the quotient 
$K^1(N)/j(K^1(M))$.

This interpretation of \pgo\ thus forces us to assert that Type IIB
RR fields on $M$ (or $N$) in the absence of branes
are classified topologically by $K^1(M)$ or
$K^1(N)$.  This extends the relation of $K$-theory to RR charges
that has been asserted in previous work.

Now we move on to the analogous situation for Type IIA.
Here, brane charge is classified by $K^1$.  More specifically, a brane
of compact support on $M$ has a charge in $K^1_{cpct}(M)$, and 
(after requiring that the equation for the RR fields has a solution)
the brane charge takes values in the kernel of
\eqn\ngo{i:K^1_{cpct}(M)\to K^1(M).}
Once again, we can express the brane charge in terms of the fields
at infinity.  The exact sequence analogous to \julpo\ is
\eqn\pulpo{\dots \to K(M)\underarrow{j} K(N)
\to K^1(M,N)\underarrow{i} K^1(M)\dots,}
where again $K^1(M,N)$ 
is the same as $K^1_{cpct}(M)$ and $j$ is the restriction
to $N$.  Hence the brane charge takes values in
\eqn\krulpo{{\rm ker}(i)=K(N)/j(K(M)).}
We interpret this to mean that for Type IIA in the absence of branes, the RR fields on $M$ (or $N$)
are classified topologically by
$K(M)$ (or $K(N)$).

Finally, for Type I, $D$-brane charge is classified by $KO(M)$ with a 
compact support condition.  Reasoning along the above lines leads to the
conclusion that RR fields on $X=\R\times M$ are classified by $KO^{-1}(M)$.

The arguments we have given are slightly formal.  But our conclusion
that in the absence of branes
 the RR fields  on a spacetime $X=\R\times M$
are classified in this way  by $K$-theory
 seems to be the only
reasonable way to reconcile the interpretation of brane charge in $K$-theory
with the fact that in gauge theory, one expects to be able to measure
the charges in terms of the fields at infinity.
Since including the first factor in $X=\R\times M$ does not change the 
$K$-groups, we can equally well say that for $X$ of the form $\R\times M$,
the RR fields are classified (for Type IIA, Type IIB, and Type I, 
respectively) by $K(X)$, $K^1(X)$, and $KO^{-1}(X)$.

Finally, we will take the additional leap of assuming that this result
is  not special to the case of $X=\R\times  M$ that we have used
to motivate the discussion.  We will assume that (in the absence of
branes) the RR fields
on an arbitrary spacetime $X$, not necessarily of the form $\R\times M$,
are classified topologically by $K(X)$ or $K^1(X)$ or $KO^{-1}(X)$.

\bigskip\noindent{\it Relation Of $K(X)$ To RR Fields}

This last statement raises the following question, which we will
consider first for Type IIA.  Given an element $x\in K(X)$ that determines
an RR field $G$,
what is the de Rham cohomology class of $G$?

\nref\aaa{E. Gava, K. S. Narain, and M. H. Sarmadi, ``On The Bound States
Of $p$- And $(p+2)$-Branes,'' Nucl. PHys. {\bf B504} (1997) 214, 
hep-th/9704006.  }%
\nref\bbb{   A. Sen, ``Tachyon Condensation On The Brane-Antibrane System,''
hep-th/9805170.}%
To answer this question, we go back to the case that $X= \R\times M$.
We consider a collection of 8-branes and $\bar 8$-branes
with world-volume $p\times M$, with $p$ being
a point in $\R$, and with arbitrary Chan-Paton bundles $(E,F)$.
Such configurations are classified topologically (modulo brane
creation and annihilation \refs{\aaa,\bbb}) by the $K$-theory
class $x$ of the pair $(E,F)$.  

\def\ch{{\rm ch}}
In crossing the branes, the de Rham cohomology class of the RR fields
jumps.  The jump is determined by the couplings of the RR fields
to the brane.  The relevant couplings were determined in \ghm.
They are usually expressed as electric couplings for the total
RR potential $C=C_1+C_3+\dots$.  The couplings are
\eqn\pkopo{\int_{p\times M}C\wedge \sqrt{\hat A}\left(\ch(E)-\ch(F)\right),}
where $\ch$ is the Chern character. 
Here $C_{2p-1}$ is the potential for the RR field $G_{2p}$.
(An additional term is needed \ref\townsend{M. B. Green, C. M. Hull,
 and P. K. Townsend,
``$D$-Brane Wess-Zumino Actions, $T$ Duality, and the Cosmological
Constant,'' Phys. Lett. {\bf B382} (1996) 65, hep-th/9604110.}
to include the electric coupling of $G_0$, since there is no $-1$-form
potential of $G_0$ in any standard sense; the effect of the addition
is summarized in eqn. (2.17) below.)
A minus sign multiplies the second term of this expression
because $\bar 8$-branes have opposite sign couplings
from $8$-branes.
Because of this minus sign and the fact that $E\to \ch(E)$ is a linear
map from bundles to cohomology classes, it follows that the coupling
on the right hand side in \pkopo\ depends only on the $K$-theory
class $x$ of the pair $(E,F)$.
 We write the difference
$\ch(E)-\ch(F)$ more succinctly as $\ch(x)$, so we can rewrite \pkopo:
\eqn\pokopo{\int_{p\times M}C\wedge \sqrt{\hat A}\,\ch(x).}
In the presence of this coupling, the equation of motion of the RR field
becomes
\eqn\lopo{d(*G)=2\pi \delta(p) \sqrt{\hat A}\,\,\ch(x),}  
where $\delta(p)$ is a delta function supported on $p\times M$.
This equation implies that $G$ jumps in crossing the brane.  If we write
$G_L$ and $G_R$ for the $G$-fields to the left and right of the brane,
the jump is given by
\eqn\mopo{*G_R-*G_L = 2\pi\sqrt{\hat A}\,\,\ch(x).}
(Both $X$ and $M$ are oriented, so there is a natural left and right.)

Had we made a duality transformation, replacing $G$ by $*G$, then 
\pkopo\ would be replaced by a magnetic coupling to the branes,
which contributes to the Bianchi identity.  With the magnetic coupling
included, the Bianchi identity becomes
\eqn\lopop{dG=2\pi\delta(p) \sqrt{\hat A}\,\,\ch(x),}
and this implies a jump in $G$ of the form
\eqn\mopom{G_R-G_L = 2\pi\sqrt{\hat A}\,\,\ch(x).}
The magnetic coupling is more in the spirit of the derivation
we gave above (in which branes were introduced as magnetic sources),
and we will take \mopom\ as the basic relation. 

Of course, since $G$ is supposed to be self-dual and the right hand side
of \mopo\ or \mopom\ is not self-dual, one should wonder what
these equations mean.  The most straightforward approach 
is to use self-duality to eliminate (for example)
$G_6$, $G_8$, and $G_{10}$,
treating the independent fields as $G_0$, $G_2$, and $G_4$.  For
$G_0$, $G_2$, and $G_4$ one would use (say) the magnetic coupling in
\mopom, while the magnetic coupling of $G_6$, $G_8$, and $G_{10}$
is included via an electric coupling of the potentials $C_{2p-1}$, $p\leq 2$,
as in \pkopo.
In section 3, we will follow a more general formalism in which one can take
an arbitrary ``mutually commuting'' set of RR periods as  independent
variables.  In that framework, if $M$ is a codimension 1 submanifold
of a spacetime $X$, the restriction of a class in $K(X)$ to $K(M)$ 
is a set of commuting data that can be treated classically and is unconstrained
by self-duality.   In this framework, we can interpret
\mopom\ as a formula for the jumps of the cohomology classes of the
$G_p$'s for all $p$ in crossing
the brane.   

Now we can finally make our proposal for the RR field determined
by a $K$-theory class.
Suppose  that in this situation, the RR-fields are classified
to the left of the branes by a $K$-theory class $a$, and to the right
of the branes by a $K$-theory class $b=a+x$.   \mopom\
expresses the difference between the $G$ field on the left and on
the right.  Let us assume that if $a=0$ then $G$ is zero (in de Rham cohomology)
to the left of the brane.
Then \mopom\ determines $G$ to the right of the brane, and we interpret
this as the RR field of the $K$-theory class $b=x$: 
 \eqn\nopo{{G(x)\over 2\pi} = \sqrt{\hat A}\,\,\ch(x).}
We will assume that this formula holds for arbitrary spacetimes
$X$ in the absence of branes, 
and not just in the situation that we have used to motivate it.

We also need the Type IIB analog of \nopo.  We propose that this
is given by the same formula, but interpreted as follows. The RR
fields of Type IIB, in the absence of branes, are determined by an element
$x\in K^1(X)$.  $K^1(X)$ is the same as $\tilde K(\S^1\times X)$
(the subgroup of $K(\S^1\times X)$ consisting of elements that are trivial
if restricted to $q\times X$ for $q$ a point in $\S^1$).  The
Chern character $\ch(x)$ is hence an element of the even-dimensional
cohomology of $\S^1\times X$.  Upon integration over $\S^1$, it maps
to an element of the odd-dimensional cohomology of $x$.  We will
abbreviate this element as $\ch(x)$ (not showing the integration over $\S^1$
in the notation).  With this notational understanding, we propose \nopo\ for
both Type IIA and Type IIB.  Given the Type IIA result, and assuming
that the Type IIB formula should have the same general form,
this is the unique formula for Type IIB 
that is consistent with the requirement of
$T$-duality between Type IIA and Type IIB in case $X=\S^1\times Y$ 
for some $Y$.

For Type I, we again propose that the RR fields are determined for
$x\in KO^{-1}(X)$ by the same formula \nopo, interpreted along the
lines suggested in the last paragraph.

\bigskip\noindent{\it $K$-Theory And Unbroken Symmetries}

As well as classifying the RR fields, we also want to classify their
symmetries.  We begin again by recalling the case of ordinary
$p$-form fields.  A potential $C_{p-1}$ with curvature $G_p$ has
a gauge-invariance $C_{p-1}\to C_{p-1}+dB_{p-2}$, with $B_{p-2}$
a two-form gauge-parameter.  An unbroken gauge symmetry is a $B_{p-2}$
such that $dB_{p-2}=0$; they should be classified mod
$B_{p-2}\to B_{p-2}+d
a_{p-3}$ (for any $(p-3)$-form $a_{p-3}$)
and modulo $2\pi$ shifts in the periods of $B_{p-2}$.  The
group of unbroken gauge symmetries is $H^{p-2}(X;U(1))$.  
This group is not necessarily connected.  By considering the long
exact sequence in cohomology derived from the coefficient sequence
\eqn\klip{0\to \Z\to \R\to U(1)\to 0,}
one can show that its group of components $\overline H^{p-2}(X;U(1))$
is the same as the torsion subgroup $H^{p-1}(X;\Z)_{tors}$ of
$H^{p-1}(X;\Z)$.

The analog for Type IIA is that, while the RR fields are classified
by $K(X)$, the unbroken gauge symmetries are classified by $K^{-2}(X;U(1))$
(which by periodicity is the same as $K(X;U(1))$), and its group of
components is the torsion subgroup
$K^{1}(X)_{tors}$.  In Fadde'ev-Popov
gauge fixing, we will have to divide by the order of this group.

For Type IIB, the group of components of the unbroken RR gauge symmetry
group is $K(X)_{tors}$, and for Type I it is $KO^{-2}(X)_{tors}$.

\bigskip\noindent{\it $K$-Theory And Cohomology}

  $G/2\pi$ as
determined in \nopo\ is not an integral cohomology class, but
(because of fractions in the power series expansion of $\sqrt{\hat A}$
and $\ch$) a rational one.  We will interpret the RR fields $G_p$ simply
as $p$-forms, with no integral structure and no attempt to define
the ``torsion part'' of $G_p$.  The integral structure is defined
at the level of $K$-theory, and in particular the torsion for RR fields
of Type IIA (in the absence of branes) is simply the torsion subgroup of 
$K(X)$.  

It is illuminating to consider briefly some examples of how
the passage from cohomology to $K$-theory mixes the RR forms.  (The examples
that follow are not used in the rest of the paper.) If
for simplicity we consider a situation in which
$\hat A=1$ (getting rid of the most complicated
fractions) and $G_0=G_2=0$, then \nopo\ gives\foot{In terms of the
formal roots $x_i$ of the Chern polynomial, $\ch(x)=\sum_ie^{x_i}$.
As we assume $G_2=0$, we have $\sum_ix_i=0$.  We also have
$c_2=\sum_{i<j}x_ix_j$, $c_3=\sum_{i<j<k}x_ix_jx_k$, leading after
some algebra to the following formulas.}  
\eqn\kloop{\eqalign{{G_4\over 2\pi} & = c_2(x)\cr
                     {G_6\over 2\pi} & ={ c_3(x)\over 2}.\cr}}
Thus, in this situation, $G_4/2\pi$ is integral, but $G_6/2\pi$ is in general
half-integral.  Why is this half-integrality not seen in the simplest
cases of brane physics?  The most obvious case of quantization of
$G_6$ arises in measuring the flux on an $\S^6$ that links a
$D2$-brane.  (In this case, $\hat A$, $G_0$, and $G_2$ are all zero
or irrelevant, so our simplifying assumptions are valid.)
Though $c_3(x)$ can be odd in general, it can be shown using the index
theorem for the Dirac operator that $c_3$ is even for any complex vector bundle
on $\S^6$, so in this situation $G_6/2\pi$ is integral.  
In general, $c_3(x)$ is
not even, but obeys a relation $c_3(x)\cong Sq^2(c_2(x))$ mod 2,
where $Sq^2$ is a certain cohomology operation (a Steenrod square).
In view of \kloop, this means that the half integral part of $G_6/2\pi$
is determined by $G_4$.  This correlation between $G_4$ and $G_6$
is a typical illustration of the differences
between $K$-theory and cohomology, and a fact that must be taken into
account, along with the electric-magnetic duality between $G_4 $ and $G_6$,
in any detailed investigation of their properties.  This and many
other subtleties of the RR fields, which otherwise would have to 
be described piecemeal, are summarized by deriving the RR fields from 
$K$-theory.

For another illustration of the consequences of reinterpreting
the RR fields in $K$-theory, we consider the torsion.
There is no way in general to attribute elements of the torsion
subgroup of $K(X)$, which we will call $K(X)_{tors}$,
 to cohomology classes of $X$ of a definite degree.
For example, for $X={\bf RP}^7$ (which arises in some orbifold studies
and was considered in \gukov),
the even dimensional cohomology of $X$ is the sum $\Z\oplus \Z_2\oplus \Z_2
\oplus \Z_2$, where the summands are $H^0$, $H^2$, $H^4$, and $H^6$.
However, $K(X)=\Z\oplus \Z_8$, where the summand $\Z$ corresponds to $H^0(X)$,
but the summand $\Z_8$ is the $K$-theory analog  of
$H^2$, $H^4$, and $H^6$ combined. 
So, if RR fields are  interpreted
in $K$-theory, there is no way to make sense separately of the torsion part of
$G_2$, $G_4$, and $G_6$.
The generator of the $\Z_8$ factor of $K(\RP^7)$
is $x={\cal L}-{\cal O}$, where ${\cal L}$ is a nontrivial flat
line bundle over ${\bf RP}^7$ (it exists and is unique
because $\pi_1(\RP^7)=\Z_2$)
and ${\cal O}$ is a trivial line bundle.  $c_1(x)$ is the nontrivial
element of $H^2({\bf RP}^7;\Z_2)$.  The element $2x$ of $K(\RP^7)$
has $c_1(2x)=0$, $c_2(2x)=c_1(x)^2\not= 0$, proving that $2x$ is nonzero
in $K(\RP^7)$.  The element $4x$ has all Chern classes zero, but
is nonetheless nonzero in $K(\RP^7)$. 
This example thus also shows that $K$-theory
elements cannot always be classified by their Chern classes.

\newsec{Self-Duality, Theta Functions, and $K$-Theory}

\subsec{Partition Function On A Closed Manifold}

Here we will describe how to interpret self-duality of quantum RR fields
in the light of $K$-theory.  To be more precise, in the limit (small
string coupling or large volume) in which the RR fields can be treated
as free fields, we will determine their quantum partition function on 
a compact manifold $X$, in the absence of branes.
The discussion is largely a reprise of
section 4.3 of \witten, repeated here to make this paper more readable, and
with some extra details.
For additional background about partition functions of self-dual
$p$-forms for $p>1$, the reader may consult \hns\ as well as section
3 of \witten. We will carry out the discussion for Type IIA, with
brief comments later on Type IIB and Type I.

The first step is to introduce an antisymmetric bilinear form $(~,~)$
on $K(X)$.  The definition is simply that $(x,y)$ is the index of the Dirac
operator on $X$ with values in $x\otimes \bar y$.  ($\bar y$ is obtained
from $y$ by complex conjugation of the bundles.)  Since
the dimension of $X$ is of the form $4k+2$, the index $i(w)$ of the
Dirac operator with values in a $K$-theory class $w$ obeys $i(w)=-
i(\bar w)$.  Hence $(x,y)=-(y,x)$.  Also, if $x_0$ is a torsion class, so
that $nx_0=0$ for some integer $n$,
then for any $x$, 
\eqn\oxno{(x,x_0)={1\over n}(x,nx_0)={1\over n}(x,0)=0.}
Hence, if $K(X)_{tors}$ is the torsion subgroup of $K(X)$, the bilinear
form $(~,~)$ is well-defined as a bilinear form on the lattice 
$\Lambda=K(X)/K(X)_{tors}$.
It can be shown by imitating the proof of Poincar\'e duality
given in \ref\bottu{R. Bott and L. Tu, {\it Differential Forms In
Algebraic Topology} (Springer-Verlag).} that the form $(~,~)$ is unimodular
on the lattice $\Lambda$.

The idea in quantizing the theory will be to write the partition function
as a sum over a maximal ``commuting'' subgroup of $K(X)$.  Here $x$ and $y$
are considered to commute if and only if $(x,y)=0$.
(One may suspect that one should somehow construct operators $\hat x$
and $\hat y$ that only commute under that condition.)  
In view of \oxno, every maximal commutative subgroup of $K(X)$ includes
$K(X)_{tors}$.  We can always (albeit not in a unique or natural way) split
$K(X)$ as $K(X)=\Lambda\oplus K(X)_{tors}$.  Given such a splitting,
if $\Lambda_1$ is a maximal commutative sublattice of $\Lambda$,
then a maximal commutative subgroup of $K(X)$ is $\bar\Lambda_1=
\Lambda_1\oplus K(X)_{tors}$.  Every maximal commutative subgroup
of $K(X)$ can be presented in this way.
It is convenient to select
a commuting sublattice $\Lambda_2$
of $\Lambda$ that is complementary to $\Lambda_1$
(in the sense that $\Lambda=\Lambda_1\oplus\Lambda_2$).

We also define a positive definite
metric on $\Lambda$. 
It is defined by the formula
\eqn\polbo{|x|^2=\int_X G(x)\wedge *G(x),}
where $G(x)$ 
is a sum of differential forms $G_p$ of all even $p$ defined as follows.
$G(x)$ is the unique {\it harmonic} differential form
such that in de Rham cohomology, $G(x)$ is determined from $x$ by the
formula obtained in section 2:
\eqn\olbo{{G(x)\over 2\pi}=\sqrt{\hat A}\,\ch(x).}
This metric depends on the metric on $X$, and has no particular
integrality properties.  

Its attractive property is as follows.  Let $\T$ be the torus
$K(X;\R)/\Lambda$, where $K(X;\R)=K(X)\otimes_\Z\R$.  (Thus,
$K(X;\R)$ is isomorphic to $\R^{n}$, with $n=2k$ the sum of the even Betti
numbers of $X$.)  Then the metric $|x|^2$ determines a metric $g$ on $\T$,
and the antisymmetric form $(~,~)$ determines a two-form $\omega$ on
$\T$.  The fact that $(~,~)$ is integral and unimodular means that
$\omega$ is integral and
\eqn\inoc{\int_\T{\omega^k\over k!}=1.}
$\omega$ is positive and of type (1,1) with respect to $g$,
so together $g$ and $\omega$ determine a Kahler structure on $X$.

The last ingredient one needs to set up the theory is a $\Z_2$-valued
function $\Omega$ on $K(X)$ such that for all $x,y\in K(X)$,
\eqn\ooplo{\Omega(x+y)=\Omega(x)\Omega(y)(-1)^{(x,y)}.}
A natural such function was defined in \witten\ as follows.
\foot{A cocycle somewhat like $\Omega$ shows up in construction of vertex
operator algebras from lattices, 
with the following difference.  In that case,
an $\Omega$ must be chosen, but the choice does not matter.
Here, there is a distinguished $\Omega$ associated with the physical
problem, and it is essential to find it.}
  For $x$ an element
of {\it complex} $K$-theory, $x\otimes \bar x$ is naturally
defined as an element of the {\it real} $K$-group $KO(X)$.  Now we must
use for the first time the fact that the spacetime dimension in string
theory (namely 10) is of the form $8k+2$.  (Our previous remarks
are valid
in any dimension of the form $4k+2$.)  In dimension
$8k+2$, there is a natural
mod two function on $KO(X)$, namely the mod two index of the Dirac operator,
which we will denote as $j$.  (Concretely, if $w$ is a real vector
bundle -- such as $x\otimes \bar x$ with $x$ a complex vector bundle --
then $j(w)$ is the number of positive chirality zero modes of the Dirac
operator on $X$ with values in $w$, mod 2.  $j(w)$ is independent of the metric
of $X$ but in general depends on the spin structure.  In \ref\atiyah{M. F. Atiyah,
``Riemann Surfaces And Spin Structures,'' Ann. Sci. \'Ecole Norm. Sup.
{\bf 4} (1971) 47.}, its dependence on the spin structure
was expressed in terms of a relation similar to 
\ooplo.)
In \witten, $\Omega$ was defined as
\eqn\refro{\Omega(x)=(-1)^{j(x\otimes \bar x)},}
and was shown to obey the basic identity \ooplo.

\ooplo\ together with \oxno\ implies that if $x_0$ is torsion, then
\eqn\rooplo{\Omega(x+x_0)=\Omega(x)\Omega(x_0).}
When we construct the partition function as a sum over a maximal
commuting subgroup $\bar \Lambda_1$ of $K(X)$, $\Omega(x)$ will enter
as a sign factor in the sum.
All the factors in the partition function except $\Omega(x)$ are invariant
under $x\to x+x_0$. Hence, given \rooplo,
the partition function will vanish under
$x\to x+x_0$ unless $\Omega$ is identically 1 when restricted to $K(X)_{tors}$.
This vanishing cannot be removed by inserting local operators (as such
operators
do not receive contributions from the torsion), and
must be interpreted as a kind of global anomaly, analogous to the
anomaly discussed for $M5$-branes in section 5.1 of \witten. 

We do not know whether there actually are ten-dimensional spin manifolds
$X$ such that $\Omega$ is non-trivial on $K(X)_{tors}$.  If this
does occur, the anomaly can be canceled as follows by wrapping a brane.
(This paragraph is not essential in the rest of the paper.)
On $K(X)_{tors}$, $\Omega$ is multiplicative as in \rooplo, and so is
a homomorphism from $K(X)_{tors}$ to $\Z_2\subset U(1)$.
The effect of a $K$-theory class that is torsion is purely to include
phases in the path integral for certain wrapped branes.  (Some examples
of this are discussed in detail in \ref\ugg{E. Witten, ``Baryons And
Branes In Anti De Sitter Space,'' JHEP {\bf 9807:006} (1998).}.)  This suggests that the anomalous
factor $(-1)^{h(x_0)}$ can be canceled by wrapping a brane.  To prove
that this is so, first recall from section 5.1 of \witten\
the situation in cohomology.  On a compact oriented
$d$-manifold $X$, there is a Pontryagin duality
\eqn\ujjp{H^p(X;\Z)\times H^{d-p}(X;U(1))\to U(1).}
The pairing here is given by the cup product followed by integration;
the fact that it is a Pontryagin duality can be proved by following
the proof of Poincar\'e duality given in \bottu.\foot{One must 
modify the argument in \bottu\ by replacing
${\rm Hom}(~,\R)$ by ${\rm Hom}(~,U(1))$; the argument
goes through in the same way with this modification since
${\rm Hom}(~,U(1))$, like ${\rm Hom}(~,\R)$,
 maps exact sequences to exact sequences.}
This duality induces a Pontryagin duality between the
torsion subgroup $H^p(X;\Z)_{tors}$ of $H^p(X;\Z)$, and the
group $\overline H^{d-p}(X;U(1))$ of components of $H^{d-p}(X;U(1))$:
\eqn\pkkp{H^p(X;\Z)_{tors}\times \overline H^{d-p}(X;U(1))\to U(1).}
$\bar H^{d-p}(X;U(1))$  is isomorphic
(under the Bockstein $\beta:H^{d-p}(X;U(1))\to H^{d-p+1}(X;\Z)$)
to $H^{d-p+1}(X;\Z)_{tors}$, so there is a Pontryagin duality
\eqn\njjp{H^p(X;\Z)_{tors}\times H^{d-p+1}(X;\Z)_{tors}\to U(1).}
These Poincar\'e duality statements have analogs for $K$-theory.
\foot{Indeed, the main points in the proof of Poincar\'e duality in \bottu\
are that there are Mayer-Vietoris sequences in cohomology theory and
that Poincar\'e duality holds for $\R^d$.  There are analogous
Mayer-Vietoris sequences in $K$-theory, and the duality statements
above are all true for $\R^d$.  Of course, in making such a duality
statement on $\R^d$ (or in general
on a noncompact oriented manifold) one must understand one of the
two factors, such as $H^p$ or $H^{d-p}$ in \ujjp,  to be cohomology
with compact support.}
On an oriented even-dimensional manifold $X$, the analog of \njjp\
is the existence of a Pontryagin duality
\eqn\qjjp{K(X)_{tors}\times K^1(X)_{tors}\to U(1).}
The pairing here is just the physical
coupling of a torsion RR field (an element
of $K(X)_{tors}$) to a torsion $D$-brane (which in Type IIA determines
an element of $K^1(X)_{tors}$). 
This Pontryagin duality means that the homomorphism $\Omega:
K(X)_{tors}\to \Z_2\subset U(1)$ is 
\eqn\jupo{x_0\to (-1)^{(\alpha,x_0)},}
for some two-torsion element $\alpha\in K^1(X)_{tors}$, where
in \jupo\ we write the pairing in an additive notation.
Physically, this means that the anomaly will be canceled by wrapping
a Type IIA $D$-brane that represents the class $\alpha\in K^1(X)$.  In section
2, we argued that on a compact manifold $X$, the $D$-brane charge should
vanish, but the anomaly means that actually the $D$-brane charge
should equal the two-torsion element $\alpha$.  This is analogous
to the situation in \ref\freed{D. S. Freed and E. Witten,
``Anomalies In String Theory WIth $D$-Branes,'' hep-th/9907189.}: 
classical reasoning seems to show
that the restriction of the class $[H]$ of the NS three-form field
$H$ to a $D$-brane world-volume $Q$ should vanish, but because of an anomaly,
$[H]$ should actually equal the two-torsion element $W_3(Q)$.  The analogy
is clear from section 5.1 of \witten, where the $W_3(Q)$ term shows
up from the restriction of $\Omega$ to torsion.

For the rest of this paper, we restrict ourselves to the case that
$\Omega$ is identically 1 when restricted to $K(X)_{tors}$.  In
this case, the sum over the torsion subgroup will give a factor
equal to the order of $K(X)_{tors}$, which we will denote as $N$.

Moreover, if $\Omega$ is identically 1 on $K(X)_{tors}$, then it can
be regarded in a natural fashion as a function on the lattice 
$\Lambda=K(X)/K(X)_{tors}$.  A $\Z_2$-valued function on this lattice
which obeys \ooplo\ determines (by a standard differential-geometric
construction that was reviewed in \oldwitten) a unitary line bundle
${\cal L}$ over $\T$ that has a connection with curvature form $\omega$.
As $\omega$ is of type $(1,1)$, ${\cal L}$ has a natural holomorphic
structure.  \inoc\ together with the Riemann-Roch theorem implies
that ${\cal L}$ has a unique holomorphic section $\Theta$.  

For any
decomposition $\Lambda=\Lambda_1\oplus\Lambda_2$ of $\Lambda$ in terms
of commutative sublattices $\Lambda_1$ and $\Lambda_2$, $\Lambda$
can be written as a sum over certain cosets of $\Lambda_1$.  This
was explained in section 3 of \witten.  To be more precise, 
because of the duality between $\Lambda_1$ and $\Lambda_2$, there
exists $\theta\in \Lambda_1$ such that, for $y\in \Lambda_2$,
$\Omega(y)=(-1)^{(\theta,y)}$.  The theta function is
\eqn\pkook{\Theta=\sum_{x\in \Lambda_1+\half\theta}\Omega(x-\theta/2)
\exp\left(i\pi(x,\tau x)\right),}
where $\tau$ is the period matrix of the lattice $\Lambda$ (with
respect to its decomposition as $\Lambda_1\oplus\Lambda_2$).  For the analogous
theory of self-dual $p$-forms, a fairly explicit explanation of how
the period matrix comes in is in \hns.

Once the theta function is constructed, the partition function of the RR fields
(assuming that $\Omega=1$ for torsion elements) is
\eqn\kklo{Z={\Theta\over \Delta}.}
Here $\Delta$ is a determinant of the nonzero modes and is completely
unaffected by all of the subtleties that we have discussed.
$\Delta$ would be the same if RR fields were ordinary differential forms,
not related to $K$-theory.  $\Delta$ has been treated for a self-dual
$p$ form in section 4 of \hns; for a self-dual three-form on 
$\T^6$, it has been explicitly incorporated in the computation in
\ref\dolanap{L. Dolan and C. R. Nappi, ``A Modular-Invariant Partition
Function For The Five-Brane,'' Nucl. Phys. {\bf B530} (1998) 683,
hep-th/9806016.}.  

One might expect in the numerator in \kklo\ a factor of $N$ from summing
over $K(X)_{tors}$, but this factor is canceled in the following way.
In Fadde'ev-Popov gauge fixing, one must  divide by the volume
of the unbroken gauge symmetry group, which is proportional to the number
of components of that group.  As we have argued in section 2, the
group of components of the unbroken RR gauge symmetry for Type IIA
is $K^1(X)_{tors}$.  The existence of the perfect pairing \qjjp\ means
that the order of $K^1(X)_{tors}$ is the same as the order of $K(X)_{tors}$,
so the factor in the numerator that comes from summing over torsion is
canceled by a similar factor in the denominator.  This cancellation
is invariant under $T$-duality to Type IIB, which exchanges the roles
of $K^1(X)_{tors}$ and $K(X)_{tors}$!

The formula \pkook\ shows that in writing
the partition function as a sum over a maximal commuting sublattice of
$\Lambda$, one in general has to sum over, roughly speaking,
 half-integral elements of
$K(X)$.  If one chooses to express the partition function as a sum over
certain RR fields via  \olbo, the conditions on the RR fields that
must be included in the sum are much more complicated.

For Type IIB, one repeats this analysis, using $K^1(X)$ instead of $K(X)$;
the definitions of $(~,~)$ and of $\Omega$ are given in \witten.
The analogous construction for Type I is as follows.  First we have
to define an antisymmetric bilinear form $(~,~)$ on $KO^{-1}(X)$.
We recall that $KO^{-1}(X)=\tilde{KO}(\S^1\times X)$, where $\tilde {KO}(\S^1
\times X)$ is the subgroup of $KO(\S^1\times X)$ consisting of elements
that are trivial if restricted to $p\times X$ for $p$ a point in $\S^1$.
For $x,y\in  \tilde{KO}(\S^1\times X)$, we have $x\otimes y \in 
\tilde{KO}(\S^1\times \S^1\times X)$.  We define $(x,y)$ to be one
half the index of the Dirac operator on $\S^1\times \S^1\times X$
with values in $x\otimes y$; this is an integer, as $x\otimes y$ is
real and $\S^1\times \S^1\times X$ has dimension of the form $8k+4$.
As for $\Omega(x)$, we reason as in the discussion of Type IIB in
section 4.3 of \witten.  $x\otimes x$ can be interpreted as an element
of $KR(\S^2\times X)$, where the involution used in defining $KR$ is
a reflection of one coordinate on $\S^2$.  By the periodicity theorem,
 $KR(\S^2\times X)=KO(X)$, and we define
$\Omega(x)=(-1)^{j(\pi(x)))}$, where $\pi(x)$ is the image of $x\otimes x$
in $KO(X)$ and $j(\pi(x))$ is its mod two index.

\subsec{ Extension To A Manifold With Boundary}

Now we will make an important extension that goes beyond what has
been said in \witten. The goal is to show, at least in part, that the formalism we have sketched
above respects the locality of quantum field theory.
(In what follows, we make use of a mathematical
result along the lines of    
\ref\dsfreed{D. S. Freed, ``A Gluing Law For The Index
Of The Dirac Operator,'' in {\it Global Analysis In Modern Mathematics},
ed. K. Uhlenbeck (Publish or Perish Press, 1993.)}. 
\nref\daifreed{X. Dai and D. S. Freed, ``$\eta$-Invariants And
Determinant Lines,'' J. Math. Phys. {\bf 35} (1994) 5155;
D. S. Freed, ``Determinant Line Bundles
Revisited,'' in {\it Geometry and Physics}, ed. J. A. Andersen et. al.
(Marcel Dekker, 1997) 187.}%
  See \daifreed\ for the roughly analogous
but more subtle case of the gluing behavior of the eta invariant.)

Using the mod 2 index, we have defined the factor $\Omega(x)$ on a closed
manifold $X$.  We want to extend the definition to define $\Omega(x)$
on a manifold with boundary, in such a way that if $X_1$ and $X_2$
have a common boundary component $B$ (with opposite orientations), 
and $X$ is obtained by gluing
together $X_1$ and $X_2$ along $B$, then $\Omega$ will be multiplicative
in the gluing. 
This multiplicativity means that for $x\in K(X)$,
if $x_1 $ and $x_2$ are the restrictions of $x$ to $X_1$ and $X_2$, then
\eqn\nomigo{\Omega(x)=\Omega(x_1)\Omega(x_2).}
(One can also formulate a similar gluing law for the case that
$X_1$ has two boundary components both isomorphic to $Y$, and $X$ is
obtained by gluing them together.)
However, for a manifold $X_1$ with boundary, $\Omega(x_1)$ will not
be defined simply as an element of the group $\Z_2$. 
 It will be defined
as an element of a non-trivial principal $\Z_2$ bundle ${{\cal P}}$.

\def\D{{\cal D}}
\def\Pf{{\rm Pf}}
\def\SC{{\cal S}}
\def\P{{\cal P}}
In general, let $X$ be a ten-dimensional spin manifold with a boundary
$Y$, of dimension nine.  (We do not assume that $Y$ is connected.)
If we write $y$ for the restriction of
$x$ to $Y$, then $y\otimes \bar y$ is a real vector bundle.
In dimension $8k+1$, the gamma matrices are real, and the Dirac
operator $\D_Y=\sum_i\Gamma^iD_i$
with values in the real bundle $y\otimes \bar y$ is a real
antisymmetric (or antihermitian) operator.  Such 
an operator in general
has a mod 2 index, but the mod 2 index of $\D_Y$ vanishes, since the mod
2 index is a bordism invariant, and $Y$ is the boundary of the spin manifold
$X$, over which $y\otimes \bar y$ extends.  
We will only consider the case that the mod 2 index is zero
on each component of $Y$; some additional subtlety is involved in extending
the discussion when this is not true.
Under this hypothesis, for a generic
metric on $Y$ and connection on $y$, $\D_Y$ has no zero eigenvalues,
and the fermion path integral for fermions on $Y$ with values in $y\otimes \bar
y$ is nonzero.
This path integral, which is often
denoted as $\sqrt{\det\,\D_Y}$, is most naturally understood as the Pfaffian
of the real antisymmetric operator $\D_Y$, so we will write
it as $\Pf(\D_Y)$.   This Pfaffian is subject to an anomaly.
 The absolute value $|\Pf(\D_Y)|$
is naturally defined as a real number (for example, using zeta function
regularization), but there can be an anomaly in the sign
of the Pfaffian.  
The most natural way to describe mathematically this sign anomaly
is to say that $\Pf(\D_Y)$
is not a real number, but takes values in a real line bundle, called
the Pfaffian line bundle.  We will denote this line bundle as $\Pf(Y)$.
The structure group of the Pfaffian line bundle $\Pf(Y)$ is
the subgroup $\{\pm 1\}$ of the real numbers (this is just a fancy
way to say that $\Pf(\D_Y)$ is well-defined up to sign); this group
is isomorphic to $\Z_2$.   So we can build a 
principal $\Z_2$ bundle
$\P(Y)$ over the  parameter space (of metrics and gauge fields on $Y$)
using the same structure functions as those of $\Pf(Y)$.
 For $x\in K(X)$, we will define $\Omega(x)$
as a section of $\P(Y)$.  Both $\Pf(Y)$ and $\P(Y)$ depend on $y$, but
we do not show this in the notation.  A fancy way to express the relation
between them is that $\Pf(Y)=\P(Y)\otimes_{\Z_2}\eta$, 
with $\eta$ a trivial real line bundle on 
which $\Z_2$ acts as the group $\{\pm 1\}$.

To facilitate the later discussion, we make a few observations
about the Dirac operator on $Y$.  The hermitian
 operator $i\D_Y$ has real eigenvalues, found by solving the
eigenvalue problem
\eqn\jutto{i\D_Y \psi =\lambda\psi,~~\lambda\in\R.}
Let $V$ be the space of all real spinor fields on $Y$ with values in 
$y\otimes \bar y$, and let $V_\C$ be the complexification of $V$.
On $V$ there is a positive definite metric
\eqn\rutto{\langle\psi_1,\psi_2\rangle=\int_Yd^9x \sqrt g (\psi_1,\psi_2),}
where the metric $(~,~)$ on the spinor fields is constructed
using a trace on $y\otimes \bar y$ and the real structure of spinors
on $Y$. We extend $\langle~,~\rangle$ to a bilinear form on
 $V_C$ by using the same
formula without any complex conjugation.  So
 $i\D_Y$ is anti-hermitian in this inner product, and
hence if $\psi_1$ and $\psi_2$ are eigenvectors of $i\D_Y$ with
eigenvalues $\lambda_1,\lambda_2$, then $\lambda_1\langle\psi_1,\psi_2\rangle
=\langle i\D_Y\psi_1,\psi_2\rangle=-\langle \psi_1,i\D_Y\psi_2\rangle
=-\lambda_2\langle\psi_1,\psi_2\rangle.$  Consequently,
\eqn\utu{\langle\psi_1,\psi_2\rangle=0~{\rm unless}~\lambda_1+\lambda_2=0.}
 For a generic metric on $Y$, the eigenvalues in \jutto\ are
all nonzero, as we have observed above.  So if we let $\SC_+$ and $\SC_-$
be the subspaces of $V_\C$ generated respectively
by the eigenvectors with positive
and negative eigenvalue, we have a decomposition
\eqn\nutto{V_\IC=\SC_+\oplus \SC_-.}  
From \utu\ it follows that $\SC_\pm$ are isotropic subspaces of $V_C$,
that is, $\langle\psi_1,\psi_2\rangle=0$ for $\psi_1,\psi_2$ both in
$\SC_+$ or both in $\SC_-$, and moreover, they are maximal isotropic
subspaces (since a vector in $\SC_-$, for example, is never orthogonal
to its complex conjugate, which is in $\SC_+$).

To define $\Omega(x)$, we introduce the Dirac operator $\D_X$ on $X$
using Atiyah-Patodi-Singer (APS) boundary conditions \ref\aps{M. F. Atiyah,
Patodi, and I. M. Singer, ``Spectral Asymmetry and Riemannian Geometry, I''
Math. Proc. Cam. Phil. Soc. (1975) {\bf 77} 43.}.  This means simply that we consider the
operator $\D_X$ to act on spinor fields on $X$ whose restriction to $Y$ is
in $\SC_-$.  For $w\in KO(X)$, we write $j(w)$ for the mod 2 index
with values in $w$, that is, $j(w)$ is the number mod 2 of positive
chirality zero modes of $\D_X$, with APS boundary conditions.
We define
\eqn\pobo{\Omega(x)=(-1)^{j(x\otimes \bar x)}.}
The right hand side is invariant under deformations of the metric
on $X$ and connection on $x$, as long as we only consider data for
which the operator $i\D_Y$ has no zero eigenvalue.  There is no natural
extension of the definition of $\SC_-$ when $i\D_Y$ develops a zero
eigenvalue, so in general $\Omega(x)$ cannot be defined continuously
as a $\Z_2$-valued function.  However, we will see later that 
\eqn\robo{\Omega(x) |\Pf(\D_Y)|}
always varies smoothly, even when one crosses the locus on which
$\Pf(\D_Y)$ develops a zero eigenvalue.  This means that $\Omega(x)
|\Pf(\D_Y)|$ can be interpreted globally as a section of $\Pf(Y)$, and
so $\Omega(x)$ is globally a section of $\P(Y)$.

Given the definition of $\Omega(x)$, we can verify the gluing law
in \nomigo.   We keep to our assumption that all boundary components,
including $B$, have zero mod 2 index.  We perform the computation using a convenient metric
and gauge connection such that the Dirac operator  $\D_B$ has no zero
eigenvalues, and the metric on $X$ looks near $B$ like $\R\times B$,
with this description being valid for a distance $t$ (in the $\R$ direction)
that is very long compared to the size of $B$.  
For $t\to\infty$, since $\D_B$ has
no zero modes, all zero modes of $\D_X$ grow or decay exponentially in the
$\R$ direction, and converge for large $t$ to zero modes on $X_1$ or on $X_2$.
Hence, the number of eigenvalues of $\D_X$ that converge to zero for $t\to
\infty$ is the sum of the number of zero modes of $\D_{X_1}$ and the
number of zero modes of $\D_{X_2}$.  This implies that the mod 2
index is additive, and gives \nomigo.\foot{If $B$ has a nonzero mod 2 index,
$\D_X$ can have a zero mode that is not localized on either side.
More care is then needed both here and in the definition of $\Omega(x)$.}

The fact that $\Omega(x)$ is an element of $\P(Y)$
has implications for the quantization of the RR fields on $Y$ in a
Hamiltonian framework.  In quantizing the zero modes of the RR fields
on $Y$, one gets a one-dimensional space ${\cal H}_y$ of
quantum states for each $y\in K(Y)$.
Superficially, it seems that ${\cal H}_y$ is canonically a copy of $\C$
-- one describes an RR quantum state by giving a number for each $y$
-- but actually ${\cal H}_y$ is isomorphic to $\P(Y)\otimes_{\Z_2}\C$.
This follows from our formalism.  Since the restriction of $x\in K(X)$ to
its boundary values $y\in K(Y)$ consists of mutually commuting
observables, the boundary values can be simultaneously specified
and treated classically.  In doing so, the factor $\Omega(x)$ is a 
factor in the path integral, and since it takes values in
$\P(Y)$, the path integral with boundary values $y$ is not a number but
an element of $\P(Y)\otimes_{\Z_2}\C$ (where again, $\P(Y)$ depends
on $y$).  Since the path integral
on a manifold with boundary should define a quantum state in the Hilbert
space of the boundary, the space of quantum ground
states for given $y$ must be
isomorphic to this.

\def\SC{{\cal S}}
\bigskip\noindent{\it Verification Of The Main Claim}

Finally, we must show that \robo\ varies smoothly as the data
on $Y$ vary.  We will explain this by analogy with a finite dimensional
situation.  Let $V$ be a real vector space of dimension $2k$ for some
$k$ with a positive definite inner product which we denote $\langle v,w\rangle$
for $v,w\in V$.  Let $V_\C$ be the complexification of $V$, to which
we extend $\langle ~ , ~\rangle $ as a bilinear form, and let
$\SC$, $\SC'$ be maximal isotropic subspaces of $V_\C$, that is, 
$k$-dimensional subspaces such that $\langle v,w\rangle = 0$ for $v,w\in
\SC$ or $v,w\in \SC'$.  Then the dimension of the intersection $\SC\cap
\SC'$ is invariant mod 2 under deformations of $\SC$ and $\SC'$ (as maximal
isotropic subspaces).  This can be proved using the one-dimensional
Dirac operator on the unit interval $I=[0,1]$ for a fermi field $\chi$
with values in $V$. In one dimension, the spin representation is 
one-dimensional, so the total number of components of $\chi$ is the
dimension $2k$ of $V$.  There is no room for curvature or holonomy, 
so we can take
the Dirac operator on the interval to be just $\D_I=d/dt$, $0\leq t\leq 1$.
 We impose boundary conditions that $\chi(0)\in \SC$
and $\chi(1)\in \SC'$.  The Dirac operator with these boundary conditions
is elliptic and skew-symmetric (here one uses that $\SC$ and $\SC'$ are
maximal isotropic), so it has a mod two index.  A zero mode is a
$t$-independent fermion with values in the intersection $\SC\cap \SC'$,
so the mod 2 index equals the dimension of this intersection mod 2, and
hence this dimension  is a topological invariant mod 2.

There actually are two connected  families of maximal isotropic subspaces.
Once we pick an orientation of $V$, they can be described as follows.
Upon picking a basis $s_1,s_2,\dots , s_k$ for $\SC$, we consider
 $\SC$ to be self-dual or anti-self-dual depending on whether
the $k$-form $ds_1\wedge ds_2\wedge \dots \wedge ds_k$ is self-dual
or anti-self-dual.  If $\SC$ and $\SC'$ are both self-dual or both
anti-selfdual, the intersection dimension is $k$ mod 2 (this is clear
upon taking $\SC=\SC'$), and if they are of opposite types, the intersection
dimension is $k-1$ mod 2.  To verify the last statement, consider the
following modification of $\SC$.  We can pick a basis such that $\bar s_1$
(as well as $s_1$) is orthogonal to $s_2,s_3,\dots , s_k$, and let
$\SC'$ be the maximal isotropic subspace spanned by $\bar s_1$ and
$s_2,s_3,\dots, s_k$.  Then $\SC'$ has opposite type from $\SC$ and
its intersection with $\SC$ has dimension $k-1$.  If two maximal isotropic
subspaces  differ in this way (by complex conjugating one basis
vector), we say they differ by an elementary modification.

\def\W{{\cal W}}
We will apply this formalism in an infinite dimensional case in which
$V$ is the space of all spinor fields on $Y$ with values in $y\otimes \bar y$.
The perturbations we consider are sufficiently soft so that the above
concepts can be applied even though $V$ is infinite dimensional.
(One apparently cannot make sense of whether $k$ is even or odd, however.)
We take as above $\SC_-$ to be the maximal isotropic subspace of $V_\C$ 
consisting of negative eigenvalues of $i\D_Y$.  We let $\W$
be the  subspace of $V_\C$ consisting of boundary values of solutions of
the Dirac
equation $\D_X\psi=0$ for $\psi$ a spinor field on $X$ valued in 
$x\otimes \bar x$.  $\W$ is isotropic since for two solutions $\psi_1$,
$\psi_2$ of the Dirac equation on $X$ (we use the same name for the
restriction to $Y$) we have
\eqn\toyo{\eqalign{
\langle\psi_1,\psi_2\rangle=&\int_Yd^9x\sqrt g(\psi_1,\psi_2)
=\int_X d^{10}x\sqrt g \partial_i(\psi_1,\Gamma^i\psi_2)\cr
=&\int_X d^{10}x\sqrt g\left((\D_X\psi_1,\psi_2)+(\psi_1,\D_X\psi_2)\right)=0.
\cr}}
General considerations about elliptic operators show that $\W$ is maximal
isotropic.  With APS boundary conditions, the space of zero modes
of $\D_X$ is the intersection $\W\cap \S_-$, so the mod 2 index is
the dimension of this intersection mod 2.

Now suppose that by varying the metric or connection on $Y$, we pass
through a locus $L $ in field space on which eigenvalues of $\D_Y$ pass
through zero.  Generically, this is where $\D_Y$ develops zero eigenvalues.
The number of such eigenvalues will be even (since the dimension
of the null space of $\D_Y$ is a topological invariant mod 2) and
generically will be precisely 2.  Let $\psi_1 $ and $\psi_2$ be the
two zero modes at some point on $L$.  Near $L$, restricted to this
two-dimensional space, $\D_Y$ (being a real, antisymmetric matrix) looks
like
\eqn\lson{\left(\matrix{0 & \epsilon\cr -\epsilon & 0 \cr}\right),}
where generically
$\epsilon$ has a simple zero on $L$ and changes sign as one crosses $L$.
The linear combination of $\psi_1 $ and $\psi_2$ that is in $\SC_-$ 
is $\psi_1+i\psi_2$ or $\psi_1-i\psi_2$ depending on the sign of 
$\epsilon$.  Hence, $\SC_-$ undergoes an elementary modification in
crossing $L$.  It follows that the mod 2 index $j(x\otimes \bar x)$
changes by $1$ in
crossing $L$, so  $\Omega(x)=(-1)^j$ 
changes sign in this crossing.  The Pfaffian $\Pf(\D_Y)$
is proportional to $\epsilon$, so its absolute value $|\Pf(\D_Y)|$ is
proportional to $|\epsilon|$ and is not smooth on $L$.  But
$\Omega(x)|\Pf(\D_Y)|\sim \Pf(\D_Y)$ varies smoothly, as we wished
to show,  since
$\Omega(x)$ changes sign precisely where $\Pf(\D_Y)$ does.

\newsec{Application To RR Periods In Type IIA}

We will now explain an illuminating application of this formalism
which gave, in fact, the original
motivation for writing the present paper.

\def\Pf{{\rm Pf}}
Let $W$ be the worldvolume of a $D$-brane in Type IIA superstring theory
on a ten-dimensional spin manifold $X$.
Its dimension is of the form $2k-1$ for some $k$.  Let $N_W$ be the
normal bundle to $W$.  $W$ is not necessarily spin
(it should be ${\rm Spin}^c$ \freed).  In
any event, because $X$ is spin, letting
 $S(W)$ denote the spin bundle of $W$ and
$S(N_W)$ denote the spin bundle of $N_W$, the tensor product $S(W)\otimes 
S(N_W)$
exists as an ordinary vector bundle.  The worldvolume fermions on $W$
takes values in this bundle.  The Dirac operator $\D_W$ is real or
pseudoreal depending on $k$, and in any event, the fermion path
integral, which is most naturally understood as a Pfaffian $\Pf(\D_W)$,
is real.  There is no problem in defining (with zeta function regularization,
for example) the absolute value $|\Pf(\D_W)|$, but the sign may have
an anomaly.  Mathematically, $\Pf(\D_W)$ is naturally defined as
a section of a real line bundle $\Pf(W)$ (the ``Pfaffian line bundle'') over
the appropriate space of fields. This line bundle may be nontrivial.

The anomaly in the fermion path integral means concretely the following.
If we go around a loop in the space of $W$'s, $\Pf(\D_W)$ might come
back with the opposite sign.  (In varying $W$, one may also vary
other data 
such as the metric 
on $X$.  To keep 
the notation simple, 
we will consider a loop of $W$'s in a fixed $X$.  
One can also let the Chan-Paton gauge
fields on $W$ vary as 
one varies $W$, but we will omit this from the notation.)   

When one goes around a loop in the space of $W$'s, $W$ sweeps out,
if things are generic enough,
a $2k$-dimensional submanifold $U\subset X$.  To keep things simple,
we will assume that this is so.  In going around the loop,
$\Pf(D)$ changes by
\eqn\pkoo{\Pf(D)\to (-1)^{\nu(U)}\Pf(D),}
where, depending on the value of $k$,
 $\nu(U)$ is the ordinary or mod 2 index of the Dirac operator $\D_W$.   
We give some details 
below on the proof of \pkoo\ and the precise index theory formula
for $\nu(U)$.

The sign factor in \pkoo\
 is the global anomaly, and when it is nonzero, it must be canceled
by the coupling of the brane to the RR fields.  The relevant
factor in the path integral (from \refs{\vafa,\ghm,\chyi,\mm}) is
\eqn\pkon{\exp\left(i\int_W 
C\wedge \sqrt{\hat A(W)/\hat A(N)}\,\ch\,(x)\right),}
with $x$ the $K$-theory class of the gauge bundle on the brane.
In going around a loop it changes by a factor
\eqn\plonk{\exp\left(i\int_U G \wedge 
\sqrt{\hat A(U)/\hat A(N)}\,\ch\,(x)\right).}
Thus, the condition that the argument of the path integral
is single-valued in going around a loop is that
\eqn\oponk{(-1)^{\nu(U)}
\exp\left(i\int_U G \wedge \sqrt{\hat A(U)/\hat A(N)}\right)=1}
whenever $U$ is the total space of a one-parameter 
family of brane world-volumes.

Note that a one-parameter family of $W$'s has topology $U=W\times \S^1$.
But our discussion below will show essentially that \oponk\ is valid
whenever a $D$-brane can be wrapped on $U$.  Global anomalies for a 
one-parameter family of $p$-dimensional objects often give a topological
restriction that is valid  in the physical problem
for a wider class of $(p+1)$-manifolds than one can get, strictly, from
a one-parameter family.  This is such a case, and indeed we will
consider below an example with $U=\S^{2k}$.
  Sometimes the extension beyond what one learns directly from global
anomalies  can be proved using conditions of locality.  Here
we will get the 
more general result by implementing the formalism of section 3. 

To try to keep things simple, we will consider a case in which
only the highest dimension RR field $G_{2k}$ contributes
(for example, because the lower $G$'s are zero).  The reason
for considering this case is that it brings out a conceptual difficulty
that we want to emphasize.  The additional contributions from the lower
$G$'s make the formulas more complicated, but do not affect this
conceptual difficulty.

If only $G_{2k}$ is relevant, the condition for anomaly cancellation is that 
\eqn\noponk{(-1)^{\nu(U)}\exp\left(i\int_U G_{2k}\right)=1.}
If $\nu$ were identically 0, this would give the naively expected
condition that the periods of $G_{2k}$ are integral multiples of $2\pi$.
More generally, if there is a differential form $\lambda$ in spacetime
such that 
\eqn\roponk{\nu(U)=\int_U\lambda}
for all $U$, then the quantization condition $G_{2k}$ is shifted
to 
\eqn\toroxo{\int_U{G_{2k}\over 2\pi}=\half\int_U\lambda +{\rm integer},}
so that it is not $G_{2k}$ but $G_{2k}+\pi\lambda$ that obeys conventional
Dirac quantization.  One can formulate a similar, though
somewhat more abstract, statement if there exists not a differential
form $\lambda$ but an element $\lambda\in H^4(X;\Z_2)$ obeying \roponk.

The case of $D2$-branes was considered in \ugwitten.  (The discussion
was actually carried out in $M$-theory and was equivalent to a Type IIA
discussion with $G_2$ and $G_0$ assumed to vanish, as we have done in obtaining
\noponk.)  In this case, a $\lambda$ with the appropriate properties
exists: it is the differential form that is related to the characteristic
class $p_1(X)/2$.

For a $D4$-brane, there is no such $\lambda$, even as an element
of $H^4(X;\Z_2)$. (The problem also has an 
analog for $D8$-branes once one allows the Chan-Paton bundle
on $W$ to vary.)
This is because there is no cohomological formula for the mod 2 index
in six dimensions.\foot{We are grateful to D. Freed and M. J. Hopkins
for explaining this to us, and to Hopkins for explaining
in detail how far one can go in the direction of such a formula and
what sort of topology is involved.}  For any given $U$, one can certainly
find a $\lambda$ such that \roponk\ is true.  But there is no $\lambda$
that works for all $U$'s.  

Thus, it seems impossible, or at least unreasonably complicated
(involving, at best, a variety of higher order cohomology operations)
 to state the appropriate quantization condition
for $G_6$ if one interprets $G_6$ as a differential form and states
the condition in terms of cohomology.  
Given the relations between RR charges and $K$-theory that
have emerged in the last few years, one wonders if a formalism
based on $K$-theory would make it easier to state the necessary
quantization conditions.   Doing this was the original goal of the present
paper.  For this, we have needed the explanation in section 2 of how
the RR fields are classified by
$K$-theory, and the material that we have surveyed
in section 3 concerning the interpretation of self-duality in the $K$-theory
language.

\bigskip\noindent{\it An Illustrative Example}

We will now make explicit what the formalism of sections 2 and 3 means
as applied to an illustrative example, and show that \noponk\ is a 
consequence.  (An extension of the same reasoning shows that
if suitably interpreted, \oponk\ is a consequence of the formalism
in sections 2 and 3.)

We want to consider a simple example in which spacetime contains
a $2k$-sphere $U$ with normal bundle $N$.  $N$ is a real vector bundle
of rank $10-2k$.  For suitable choice of $N$,
$\nu(U)$ will be nonzero; we want to explore the quantization of the
RR form $G_{2k}$ on $U$.  

We could take the spacetime manifold $X$ to be the total space of
the bundle $N$.  However, the framework of section 3 is most straightforward
for compact $X$.  Hence, we replace $N$ by a sphere bundle.  This is done
by adding a point at infinity to each fiber of $N\to U$.  The fibers
are copies of $\R^{10-2k}$; compactifying each fiber by adding a point
at infinity, we compactify the total space of $N$ to a manifold $X$
that is a sphere bundle over $U$, with fibers $\S^{10-2k}$.

A homology basis of $X$ is given by the following four classes:
a point $p\in X$; $U$, embedded in $X$ as the zero section of 
$N$; a fiber $F$ of the sphere bundle $X\to U$; and $X$ itself.
These are all spin manifolds, so a brane wrapped on any one of them
with trivial Chan-Paton bundle gives an element of $K(X)$.  We denote
the corresponding elements of $K(X)$ as $[p],[U],[F]$, and $[X]$.  Using
the Atiyah-Hirzebruch spectral sequence, it can be shown $K(X)\cong \Z^4$
with these four classes furnishing a basis.  The bilinear form
on $K(X)$ is given by $([p],[X])=1$, $([U],[F])=(-1)^k$, with other
components vanishing.  (The factor $(-1)^k$, which will not play an important
role,
comes from the complex conjugation of the second factor in the definition
of the antisymmetric form $(x,y)$ on $K(X)$.)

The $\Z_2$-valued function $\Omega$ is completely determined by
its value for the four basis elements together with the fundamental
relation \ooplo. 
We will see that if $V$ is any even-dimensional spin submanifold of $X$
and $[V]$ is the corresponding $K$-theory class, then
\eqn\nurjy{\Omega([V])=(-1)^{\nu(V)}.}
From this, it follows in the case at hand that $\Omega([V])$ is $+1$
if $V$ is $p$, $F$, or $X$, while $\Omega([U])=(-1)^{\nu(U)}$ is in general
non-trivial. ($\nu(p)$ is zero because the Dirac
index on a point, with values in an even rank bundle, is zero mod 2.
$\nu(F)$ is zero because $F$ has trivial normal bundle,
and positive scalar curvature, so the relevant Dirac operator has
no zero modes. Finally, $\nu(X)$ is zero because $X$ is a fiber bundle with fibers of
positive scalar curvature, so the Dirac equation has no zero modes.
The statements about $\nu(F)$ and $\nu(X)$ use the fact that the Chan-Paton
bundles are trivial.)

Given this, let us discuss the quantization of the RR periods.
The theta function of $K(X)$ is constructed as a sum over a maximal
commuting lattice $\Lambda_1$, which we take to be generated by $[F]$
and $[X]$.  We take the complementary lattice $\Lambda_2$ to be
generated by $[p]$ and $[U]$.  The theta function is constructed
as a sum over the coset $\half\theta+\Lambda_1$ in $\half\Lambda_1$,
where $\theta\in\Lambda_1$ is such that 
\eqn\polly{\Omega(x)=(-1)^{(\theta,x)}}
for $x\in\Lambda_2$.  In view of \nurjy\ and the structure of the bilinear form
$(~,~)$, this means that
\eqn\ploog{\theta=[F].}
Hence the theta function is constructed as a sum over the coset
\eqn\olpog{\half [F]+\Lambda_1\subset \half \Lambda_1.}
The theta function is constructed, in other words, as a sum over
elements of the form 
\eqn\oofgo{(n+\nu(U)/2)[F]+m[X]} 
with integers $n$ and $m$.

Concretely, $[F]$ corresponds to an RR form $G_{2k}$ (since
$F$ has codimension $2k$) of delta function support on $F$, such that
\eqn\profo{\int_U{G_{2k}\over 2\pi} = 1.}
Analogously, $[X]$ corresponds to an RR form $G_0$ such that
\eqn\jurofo{{G_0\over 2\pi}=1.}
\oofgo\ shows that the theta function is constructed
as a sum over elements of $\half\Lambda_1$
that correspond to RR forms with
\eqn\nurofo{\int_U{G_{2k}\over 2\pi} = {\nu(U)\over 2} ~{\rm mod}~\Z.}
In this sense, the mod 2 index shifts the quantization of the RR forms
in the expected way.

\subsec{Computation Of $\nu(U)$}

We still need to describe why the global anomaly $\nu(U)$ is
a mod 2 index, and to show that $\Omega([U])$ is determined by the
same mod 2 index.

The details of the evaluation of the global anomaly depend on the dimension
of the brane worldvolume $W$.  The two cases in which $\dim\,(W)$ is
of the form $4n-1$ are somewhat similar, so we consider them first,
followed by the two rather similar cases with $\dim\,(W)$ of the form
$4n+1$.

\bigskip\noindent{\it $W$ of Dimension $4n-1$}

If $W$ is three-dimensional (the case already considered in 
\ugwitten), then the spinors on $W$ are
pseudoreal.  The Hermitian Dirac operator $i\D_W=i\Gamma^ID_I$
on $W$ with values in any real bundle has an antiunitary
``complex conjugation'' symmetry $\tau$, with $\tau\D_W=\D_W\tau$,
and $\tau^2=-1$.\foot{In a local Lorentz frame, one can take the
gamma matrices to be the usual $2\times 2$ Pauli sigma matrices,
with $\sigma_2$ imaginary and the others real.  $\tau$ is then
$\sigma_2$ times complex conjugation; it commutes with $i\sigma^ID_I$.}  
If $\psi$ is an eigenfunction of $\D_W$, then 
$\tau\psi$ is an eigenfunction with the same eigenvalue; it cannot
be a multiple of $\psi$ since $\tau\psi = c\psi$ for complex $c$
would imply,
given the properties of $\tau$, that $\bar c c = -1$.
Hence the eigenvalues of $\D_W$ appear in pairs.
In our case, the Dirac operator $\D_W$ that we want acts on spinors with values
in $S(N_W)$, where $N_W$ is the  normal bundle  to $W$, and
$S(N_W)$ are the spinors of $N_W$.
$N_W$ has rank seven, and $S(N_W)$ is a real bundle.

Now, when we go around a loop in the space of $W$'s, eigenvalue pairs
may pass through zero.  Every time this occurs, the Pfaffian $\Pf(\D_W)$
changes sign.  So $\nu(U)=\Delta$, where $\Delta$ is the net number of
times a pair of eigenvalues passes through zero from the positive
to negative direction.  A one parameter
family of Dirac operators on $W$ glue together to make a Dirac operator
on $U=\S^1\times W$ (assuming the metric on $W$ is kept fixed in the family;
more generally, $U$ is a fiber bundle over $\S^1$ with fibers copies of $W$,
but we will not build this into the notation). 
A standard argument relating spectral flow in three dimensions
to Dirac zero modes in four dimensions \ref\apstwo{M. F. Atiyah,
V. K. Patodi, and I. M. Singer, ``Spectral Asymmetry And Riemannian Geometry. III,'' Math. Proc. Camb. Phil. Soc. (1976) {\bf 79} 71.}
shows that the index $i(S(N_W))$
of the Dirac operator with values in $S(N_W)$ is $2\Delta$, so
$\nu(U)=i(S(N_W))/2$.  This can be more conveniently written as follows.
The four-manifold $U$ has in the string theory spacetime $X$ a normal
bundle $N$ of rank six; one has $N_W=T\S^1\oplus N$ with $T\S^1$ the
tangent bundle to $\S^1$.  As $T\S^1$ is trivial,  it follows that the spin
bundle $S(N_W)$ of $N_W$ is  the same as the spin bundle $S(N)$ of $N$.  
But the description in terms of $N$ gives a simplification; as $N$ has even rank, its spin bundle has a chiral
decomposition as  $S(N)=S_+(N)\oplus S_-(N)$.  The two summands
 are related by complex
conjugation, and hence the index of the Dirac operator with values
in $S_+(N)$ equals that with values in $S_-(N)$.  So $i(S(N_W))/2
=i(S_+(N))=i(S_-(N))$.  The final result for $\nu(U)$ is
then in this
case
\eqn\torido{\nu(U)=i(S_+(N))=i(S_-(N)).}

If $W$ is seven-dimensional,
everything is the same except the details of constructing the pseudoreal
symmetry $\tau$.
One can pick seven $8\times 8$ gamma matrices $\Gamma^i$ that are
imaginary and square to $+1$, so the hermitian Dirac operator
$i\Gamma^ID_I$ on spinors of $W$ is real.  However, we want the Dirac
operator on spinors on $W$ with values in $S(N_W)$, and (as $N_W$
has rank three) the spinors of $N_W$ are pseudoreal.  Because
of the pseudoreality of $N_W$, there is again a complex conjugation
symmetry $\tau$ of $i\D_W$ with $\tau^2=-1$.\foot{If the generators
of the $SU(2)$ structure group of $N_W$ are Pauli matrices $\sigma_i$
with $\sigma_2$ imaginary and the others real, then $\tau$ is the
product of complex conjugation with $\sigma_2$.} The rest of the argument
is the same.  The eigenvalues of $i\D_W$ are paired by $\tau$, and 
in a one-parameter family of $W$'s, there is again a possibility of spectral
flow.  So again if $\Delta$ is the net number of times an eigenvalue pair
passes through zero, then the number of sign changes of $\Pf(\D_W)$ is $\nu(U)=\Delta$.  And again, $\Delta$ is
half the index of the Dirac operator on $U$ with values in $S(N_W)$.
Once again, letting $N$ be the normal bundle to $U=\S^1\times W$,
we have $S(N_W)=S_+(N)\oplus S_-(N)$, and the same reasoning leads again
to \torido.

\bigskip\noindent{\it $W$ of Dimension $4n+1$}

Now suppose that $W$ is five-dimensional.  Then (as the normal bundle
to $W$ is of rank five and spinors of $SO(5)$ are pseudoreal)
the spinors
on $W$ and the spinors on its normal bundle $N_W$ are both pseudoreal,
so the Dirac operator $\D_W$ is real.  This means that the eigenvalues
of the Hermitian operator $i\D_W$ occur in pairs with opposite sign: if 
$i\D_W\psi=\lambda\psi$, then $i\D_W\bar\psi=-\lambda\bar\psi$.
It is still true that the Pfaffian $\Pf(\D_W)$ changes sign
every time an eigenvalue pair crosses zero, but this time the eigenvalues
in the pair are crossing from opposite directions.  We let $\Delta$
be the number of eigenvalue pair crossings mod 2, so $\nu(U)=\Delta$.
(In contrast to the case where $W$ has dimension $4n-1$, the number
of eigenvalue pair crossings is only a topological invariant mod 2 since
eigenvalues are crossing in opposite directions.)

The relation between a one parameter family of Dirac operators on $W$
and a Dirac operator on $U=\S^1\times W$ is now that the mod 2 spectral
flow $\Delta$ on $W$, for spinors with values in $S(N_W)$, equals
the mod two index of the Dirac operator on $U=\S^1\times W$, with
values in the same bundle.  We recall that this mod 2 index
$j(S(N_W))$ is defined as the number of positive chirality zero
modes of the Dirac operator on $U$ with values in $N_W$ (regarded as a bundle
on $U$), mod 2.  From $N_W=T\S^1\oplus N$, we again have $S(N_W)
=S(N)=S_+(N)\oplus S_-(N)$, so $j(S(N_W))=j(S_+(N))+j(S_-(N))$. But the
two terms on the right are in general not equal, unlike the case
when $W$ has dimension $4n-1$.  So our result is
now
\eqn\kko{\nu(U)=j(S(N))=j(S_+(N))+j(S_-(N)).}

For $W$ nine-dimensional, the analysis is much the same.
One change is  the explanation
of why $\D_W$ is a real operator.  $SO(9)$ and $SO(1)$
both have real spin representations, so the spinors of $W$ and of
its normal bundle are both real, and hence $\D_W$ is real.

The other change is that, 
since $SO(1)$ is trivial, the spinors of the normal bundle
are a trivial rank one real bundle.  So to get an anomaly, we must
let either the Chan-Paton bundle on $W$ or the metric on $W$ vary.
Also, if we want to think of the total space $U=\S^1\times W$ of a family
of $W$'s as a submanifold of the spacetime $X$ (we could consider more
general cases if we adopt 
a somewhat more abstract notation), we must for dimensional
reasons take $U=X$.  As a result, the normal bundle $N$ to $U$ in $X$ is
of rank zero, and the notation in \kko\ needs some clarification.
As there are no gamma matrices, we consider the Clifford algebra
of a rank zero vector space to consist only of scalars; there is only
one irreducible representation, of dimension 1, so $S_+(N)$ is a trivial
one dimensional bundle and $S_-(N)$ is empty.  With this interpretation,
\kko\ can be justified by the same arguments, but is perhaps more
clearly written as
\eqn\koko{\nu(X)=j,}
where 
$X$ is endowed with a set of space-filling branes carrying a Chan-Paton
bundle with $K$-theory class $x$, and $j$ is the mod two index of the
Dirac operator $\D_X$ on spinors with values in $x\otimes \bar x$.
The justification for \koko\ is the same that we gave for $W$ of dimension
five: the number of sign changes of $\Pf(\D_W)$ in a one parameter
family equals the number $\Delta$ of level crossings, mod 2;
and this in turn equals the mod 2 index of the Dirac operator,
in this case on $U=X$.

\bigskip\noindent{\it Summary}

The main difference between these various cases
is that when $W$ has dimension $4n-1$, the anomaly is given by
an ordinary index that can be  computed using a differential form.
This leads to the type of description given in \ugwitten\ for $W$ of
dimension three -- a shifted quantization condition that can be expressed
in terms of differential forms.  However, for $W$ of dimension $4n+1$,
we run into a mod 2 index that cannot be described cohomologically.
To make sense of the anomaly in these cases, the reinterpretation of the
RR fields in $K$-theory is extremely useful.

The above formulas for $\nu(U)$ can be stated in a more unified way.
For any vector bundle $T$ over $U$, let $n_+(T)$ and $n_-(T)$ be 
the numbers
of positive and negative chirality zero modes of the Dirac operator on $U$
with values in $T$.  Then in all cases we have
\eqn\mxin{\nu(U)=n_+(S_+(N))-n_+(S_-(N))~{\rm mod}~2.} 
For $U$ of dimension $4n+2$, this is equivalent to \kko, since
$j(S_\pm(N))=n_+(S_\pm(N))$ mod 2.  For $U$ of dimension $4n$,
it is equivalent to \torido, since $i(S_+(N))=n_+(S_+(N))
-n_-(S_+(N))$, and by complex conjugation $n_-(S_+(N))=n_+(S_-(N))$.
\mxin\ is a convenient expression for comparison with the computation that
we are about to perform.

\subsec{Computation of $\Omega$}

Finally, we want to show that for any even-dimensional spin submanifold
$V$ of $X$, $\Omega([V])=(-1)^{\nu(V)}$.

Let $N$ be the normal bundle of $V $ and $S_\pm(N)$ the associated
spin bundles.  The class $x=[V]$ is the $K$-theory class $(S_+(N),S_-(N))$
where $S_\pm(N)$      are understood as bundles on $X$ in the following
sense.  First, let $R$ be a tubular neighborhood of $V$ in $X$.
Then there is a projection $\pi:R\to V$, and one pulls back $S_+$, $S_-$
to bundles on $R$ that we will denote by the same names.  Then, away
from the zero section of $R$, one has an isomorphism $T:S_+\leftrightarrow S_-$
via the usual tachyon field $T=\vec\gamma\cdot \vec\phi$, where $\vec\gamma$
are gamma matrices on $S_+(N)\oplus S_-(N)$ (as usual, the gamma matrices
reverse chirality and exchange the two factors), and $\vec\phi$ are
coordinates in the normal direction.  Using this isomorphism,
$x=(S_+,S_-)$ can be understood as a $K$-theory class on $R$ that is
trivial away from $V$ and hence (maintaining this triviality away from
$R$) can be extended over $X$.  Concretely, after perhaps replacing
$(S_+(N),S_-(N))$ by $(S_+(N)\oplus F,S_-(N)\oplus F)$ for some $F$,
one can extend $S_\pm(N)\oplus F$ over $X$ such that the  tachyon
field defined in $R$ by $T=\vec\gamma\cdot \vec\phi\oplus 1$ extends
over $X$ and is an isomorphism away from $V$.

Before attempting to compute the mod 2 index with values in $x\otimes
\bar x$, we consider a slightly simpler problem.
Suppose that we want to compute the index of the Dirac operator on $X$
with values in the $K$-theory class $x$.  
The result we want to justify is that the Dirac operator           on
 $X$, for spinors with
values in $x$, has the same index as the Dirac operator on $V$
for spinors with values in a trivial line bundle.
One way to do the computation is
 to consider the Dirac operator $i\D_X$ on $X$, acting on
$S_+(N)\oplus S_-(N)$, or possibly $(S_+(N)\oplus F)\oplus (S_-(N)
\oplus F)$.  (In the computation we are about to perform, $F$ is
irrelevant, as we will see momentarily).
We perturb this operator 
  to $i\tilde D_X=i\D_X+w       T$,
where $T$, which exchanges $(S_+(N)\oplus F)$ with $(S_-(N)\oplus F)$,
is the tachyon field constructed in the last paragraph, and $w      $ is
 a real number that varies from 0 to infinity.  For $w      =0$,
$i\tilde D_X=i\D_X$, and for $w      \to\infty$, the fermions
are everywhere very massive, except near $V$, where the mass term
of the fermions with values in $S_\pm(N)$ (but not those with values in $F$)
vanishes.  Hence, eigenstates of $i\tilde D_X$ whose eigenvalue is small
for $w      \to\infty$ are localized near $N$, and are sections of
$S_+(N)\oplus S_-(N)$ -- the details of the choice of $F$ and the
extension of the bundles over $X$ are irrelevant.  The eigenvalue
problem $i\tilde D_X\Psi=\lambda \Psi$ is solved, for large $w$ and
small $\lambda$, by a kind of Born-Oppenheimer approximation.
First one solves the Dirac equation in the normal
directions.  This equation has a unique zero mode $\Psi_0$ -- this
is the basic local fact used in building $p$-branes as bound states
of $(p+2k)$-branes for arbitrary $k$ \refs{\aaa,\bbb }.  Then one
solves the Dirac equation on $V$ with an ansatz $\Psi=\Psi_V\otimes \Psi_0$,
with $\Psi_V$ being a spinor field on $V$.
Then $\Psi_V$ obeys an ordinary Dirac equation on $V$ (with values in
a trivial bundle as the line bundle generated by the $\Psi_0$'s is trivial).
So the low-lying spectrum of $i\tilde D_X$ converges, for $w\to\infty$,
to the spectrum of $i\D_V$, and the two operators have the same
index.

For our present purposes, we want not the ordinary index with values
in $x$ but the mod two index with values in $x\otimes \bar x$.
First of all, with $x=(S_+(N),S_-(N))$, $\bar x$ is equal to
either $x$ or $-x$ depending on whether complex conjugation maps
$S_\pm(N)$ to themselves or exchanges them.  This depends on the 
rank of the normal bundle.  In any event, the minus sign will not affect
the mod 2 index.  Now we need to compute the number of positive
chirality zero modes mod 2 of the Dirac operator $i\D_X$ acting
on $x\otimes \bar x$.  For this, we again introduce a tachyon
perturbation, and replace $i\D_X$ by $i\D_X+wT$.  It is up to
us to pick a convenient $T$, and we pick $T$ to be the same
tachyon field used in the last paragraph, acting on $x$ alone --
we do not include any tachyon field acting on $\bar x$.  The localization
argument for large $w$
replaces the factor of $x$ in $x\otimes \bar x$ with
the unique zero mode $\Psi_0$ in the normal direction.  So finally
we reduce to a Dirac operator $i\D_V$ on $V$ with values in
$\bar x=\pm(S_+(N),S_-(N))$.  Hence the number of zero eigenvalues
of $i\D_X$ on spinors with values in $x\otimes \bar x$ is
$n_+(S_+(N))-n_+(S_-(N))$ mod 2.  Comparing with \mxin, we see that this
statement is equivalent to the claim \nurjy\ that we have aimed to justify.

\subsec{Some Remarks On $D$-Brane Global Anomalies}

We have thus seen that by classifying RR fields by $K$-theory
and properly interpreting self-duality in the quantum theory,
we get, without looking at $D$-branes,
 the  results that would be expected to follow from
Type IIA $D$-brane global anomalies.
It would be nice also to  look at the $D$-branes 
and show that their global world-volume
anomalies cancel.  To do this effectively, one would like to have
a natural way to describe couplings of $D$-branes to RR fields in the
$K$-theory language.  Not having this, we will content ourselves
with looking at a special case.

Before analyzing the special case, we will try to describe its
theoretical significance.  On a closed manifold $X$, the total $D$-brane
charge must vanish (assuming the torsion anomaly described in section 3
does not come into play), so if we had a formalism in which it was manifest
that the anomalies depend only on the total $D$-brane charge, there
would be nothing to prove: the $D$-branes could not contribute to the
anomalies, and  the discussion would reduce to the formalism
we have presented.

Even if $X$ has a nonempty boundary $Y$, we can use the reasoning
of section 2.  The total $D$-brane charge in Type IIA is measured by
the RR fields on $Y$, which are classified by a class in $K(Y)$.
Near infinity we suppose that $X$ looks like $\R\times Y$.  Consider
a collection of $D8$-branes and $D\bar 8$-branes supported on $p\times Y$,
where $p$ is a point in $Y$.  Suppose that in the interior of $X$ (in the
compact part of $X$ that is bounded by $p\times Y$)
the RR fields vanish.   The $D8$-$D\bar 8$ configuration is classified
by an element $y\in K(p\times Y)=K(Y)$.  In crossing the branes,
the RR fields ``jump'' by $y$, so as we assume they vanish in the interior
of $X$,
they are classified at infinity by the element $y$ of $K(Y)$.

If we had a framework in which the anomalies manifestly depend
only on the $K$-theory classes of the branes, the above example would
be ``universal,'' as it enables us to get any desired set of RR fields
at infinity, and any set of branes that produce the same fields at infinity
are in the same topological class.  
Even though we do not have a formalism with the requisite
properties, it is still instructive to examine this example.

We can simplify the discussion further.  We will assume that the theory
has a reasonable degree of locality so that we can ``cut and paste.''
Using this, we can reduce to the following simple situation.  We let $q$ be
a point on $\R$ to the ``interior'' of $p$, so that cutting
$X$ on $q\times Y$ splits it into two pieces $X_1\cup X_2$
with the following
properties.  $X_1$ is  equivalent topologically to $X$ and
contains no branes, and $X_2$ contains the branes.  In fact, $X_2$
is a copy of 
$\R^+\times Y$, and the $D8-D\bar 8$ system is wrapped on $p\times Y$ for
some $p\in \R^+$. $q$ corresponds to the boundary $0$ of $\R^+$, and
the other end of $\R^+$ at infinity corresponds to the original boundary
of $X$.  We write $Y_L$ and $Y_R$ for the two ends of $X_2$, roughly
$q\times Y$ and $\infty\times Y$.
 $X_1$ and $X_2$ can be glued on their common boundary
$Y_L=q\times Y$ to make $X$. The path integral on $X$ is a product of
path integrals on $X_1$ and $X_2$ with a sum over physical states on the
common boundary.

As $X_1$ contains no branes, the path integral on $X_1$ is governed
by the formalism of sections 2 and 3.  We gave the definition of the
function $\Omega$ for a manifold with boundary
at the end of section 3. 

The new ingredient is $X_2$, which does contain branes, and has
the two boundary components $Y_L$ and $Y_R$.
   On $Y_L$
the RR fields are trivial, and on $Y_R$ they are controlled by the
$K$-theory class $y$.  According to the discussion in section 3,
the RR path integral on $X_2$ should give not a number but 
a section of $\Pf(Y_1)\otimes \Pf(Y_2)$, where $\Pf(Y_1)$ is the
Pfaffian line of the Dirac operator on $Y_1$ with values in the $K$-theory
class $0\otimes \bar 0 = 0$,
and $\Pf(Y_2)$ is the Pfaffian line of the Dirac operator on $Y_2$
with values in the $K$-theory class $y\otimes \bar y$.  
Actually, the Pfaffian line of the Dirac operator
with values in the $K$-theory class 0 is canonically trivial, so
the path integral should take values in the Pfaffian line $\Pf(Y_2)$,
or simply (as $Y_2$ is isomorphic to $Y$) $\Pf(Y)$, where as in section 3,
the $K$-theory
class $y\otimes \bar y$ is understood in the definition of $\Pf(Y)$.

What about the brane anomaly?  The $D8-D\bar 8$ system contains
worldvolume fermions in the adjoint representation, that is,
with values in $y\otimes \bar y$.  So the brane path integral
is a section of the very same Pfaffian line $\Pf(Y)$ that we have just
met.   All is in order.
There is no need to look for any additional anomaly cancellation mechanism.
And that is just as well, since, there being
no cohomological formula for the global holonomy of $\Pf(Y)$, this would
be an exceedingly difficult anomaly to cancel in a more conventional way.

\bigskip
Work of GM was supported in part by DOE grant DE-FG-02-92ER40704.
GM also thanks the Institute for Advanced Study for hospitality and the
Monell Foundation for support during the initiation of this work.
Work of EW was supported in part by NSF Grant PHY-9513835 and the
Caltech Discovery Fund.  We are grateful to D. Freed and M. J. Hopkins
for encouraging us to express everything in terms of $K$-theory,
and to M. F. Atiyah, D. Freed,
M. J. Hopkins, G. Segal, and I. Singer
for detailed explanations of a variety of mathematical
points.  We also thank E. Diaconescu for discussions  about RR fields.
\listrefs
\end